\documentclass[twocolumn,aps,prb,epsf,graphics,psfig]{revtex4}
\usepackage{graphicx,graphics} 

\def\s{\sigma}
\def\ss{\Sigma}
\def\up{\uparrow}
\def\dd{\downarrow}
\def\las{\langle}
\def\ras{\rangle}
\def\la{\left\las}
\def\ra{\right\ras}
\def\nn{\nonumber}
\def\kp{{\bf k_\parallel}}

\begin{document}
\title{Temperature dependence of surface magnetization in local-moment systems}
\author{Alireza Saffarzadeh}
\altaffiliation{E-mail: a-saffar@tehran.pnu.ac.ir}
\affiliation{$^1$Department of Physics,
Payame Noor University, Nejatollahi St., 159995-7613 Tehran, Iran \\
$^2$Computational Physical Sciences Research Laboratory,
Department of Nano-Science, Institute for studies in theoretical
Physics and Mathematics (IPM), P.O. Box 19395-5531, Tehran, Iran}
\date{\today}
\begin{abstract}
We present a theory to study the temperature-dependent behavior
of surface states in a ferromagnetic semi-infinite crystal. Our
approach is based on the single-site approximation for the
\emph{s-f} model. The effect of the semi-infinite nature of the
crystal is taken into account by a localized perturbation method.
Using the mean-field theory for the layer-dependent
magnetization, the local density of states and the electron-spin
polarization are investigated at different temperatures for
ordinary and surface transition cases. The results show that the
surface magnetic properties may differ strongly from those in the
bulk and the coupling constant of atoms plays a decisive role in
the degree of spin polarization. In particular, for the case in
which the exchange coupling constant on the surface and between
atoms in the first and second layer is higher than the
corresponding in the bulk, an enhancement of surface Curie
temperature and hence the spin polarization can be obtained.
\end{abstract}
\maketitle
\section{\bf Introduction}
Knowledge of the spin-dependent electronic structure at surfaces
and interfaces plays an increasingly important role when
assessing possible use of novel magnetic materials for spintronic
applications \cite{Labella}. A surface breaks the translation
symmetry of a system and changes local quantities such as local
density of states (LDOS) which is significantly different from
the bulk density of states (DOS). Experimental study on magnetic
surfaces has shown that the order-disorder phase transitions are
affected by the surfaces and may differ markedly from that in the
bulk \cite{Koon,Stam,Liu1,Durr,Liu2}. Indeed, the atoms at the
surface layer have fewer neighboring atoms in comparison with the
bulk, and it is well know that the magnetic order is strongly
modified due to the lower coordination number. Moreover, the
changes of the lattice constant near the magnetic surface also
modify the exchange interaction constants. Therefore, the reduced
translational symmetry may decrease or increase the magnetic
stability.

Depending on the value of exchange coupling constant, two
different behaviors at the surface can be obtained: (i) If the
coupling constant is smaller than a critical valve, the bulk and
surface ordering occur at the same temperature. This case is
called \emph{ordinary transition}. Experiments at the bulk Curie
temperature have shown that in general case, the magnetization of
thin films and surfaces is lower than the bulk one, and as the
thickness of the films decreases, the magnetization also
decreases. Ni and Fe correspond to this kind of systems
\cite{Celotta,Alvarado}. (ii) If the exchange coupling constant is
larger than a critical value, which is itself larger than the bulk
coupling constant, the Curie temperature of thin films and
surfaces is higher than the bulk Curie temperature. This case is
called \emph{surface transition} where the surface layer alone
orders, while the bulk remains disordered. This remarkable
phenomenon was observed for various compounds, such as for example
Gd \cite{Weller85,Rau87}, Tb \cite{Rau88} and NiO
\cite{Mary,Bata}. However, for Gd it should be mentioned that
there are also newer studies claiming that no separate transition
at the surface exists (e.g. see Refs. \cite{Donath,Arnold}).
Challenges in experimental studies of surface magnetism stem from
the difficulties of preparing and characterizing clean,
structurally and magnetically ordered Gd surfaces.

From theoretical point of view, the spin-dependent band structure
of films and surfaces have been investigated in local-moment
systems in which the magnetic moment originates from a partially
filled shell of the atoms and being strictly localized at each
magnetic ion site. Kurz \emph{et al.} \cite{Kurz} studied the
magnetism and the electronic structure of Gd(0001) surface on the
basis of density functional theory. Schiller \emph{et al.}
\cite{Schiller1,Schiller2} investigated the temperature-dependent
band structure of ferromagnetic semiconductor films and surfaces
in the \emph{s-f} model \cite{Nagaev}. Using a moment-conserving
decoupling procedure for suitable defined Green's functions, the
layer-dependent magnetizations and the LDOS were studied at
different temperatures.

The aim of this work is to present the single-site coherent
potential approximation (CPA) for the \emph{s-f} to investigate
the spin-polarized surface states in local-moment systems. To our
knowledge, this is the first detailed study of the surface
effects on electron-spin polarization using the CPA. In this
work, based on the single-site CPA for the \emph{s-f} model, we
intend to develop a theory for electronic states of a magnetic
semi-infinite crystal, which is applicable in wide range of
temperature and the \emph{s-f} exchange coupling strength.

The CPA is available when the scattering by the \emph{s-f}
exchange interaction is equivalent to the scattering by a
completely random short-ranged potential. In this approximation
the multiple scattering on the single site is taken into
consideration. The single-site CPA gives reasonable results for
both weak and strong exchange interaction limits, and hence is
fairly good for any values of bandwidth, scattering potential and
temperature \cite{Taka96}. The single-site CPA, however, does not
consider the exchange scattering caused by the \emph{f}-spin
correlation between different sites, which plays an important role
around Curie temperature in local moment systems. In fact, the
collective mode, correlation, and/or clustering effect of
localized spins are completely beyond the scope of the single-site
CPA.

The crystal is assumed to consist of completely filled and well
ordered ferromagnetic layers and its surface is flat. Thus, the
effects due to vacancies, islands, steps or intermixings on the
magnetic properties are not considered here. However, the effect
of magnetic anisotropy will be taken into account in the values of
coupling constant between atoms on the surface of crystal.

The article is organized as follows: In section II, by applying
the single-site approximation to the \emph{s-f} model, we present
a theory to study the surface effects on the spin-dependent
electronic states in a magnetic semi-infinite crystal. In
appendix A, we obtain the Green's function for a nonmagnetic
semi-infinite crystal and in appendix B, a procedure is given for
calculation of the layer-dependent magnetization in the frame
work of the mean-field approximation . The results of numerical
calculations of the self-consistent equations for the LDOS and
the electron-spin polarization are discussed in section III for
two cases: ordinary and surface transitions. In section IV, we
conclude the article with a summary.

\section{\bf Model and formalism}
We consider a magnetic semi-infinite crystal which can be
described by a single-orbital tight-binding Hamiltonian with
nearest-neighbor hopping $t$ on a simple cubic (s.c.) crystalline
lattice with lattice constant $a$. The structure is obtained by
stacking layers parallel to the free surface and we choose the
(001) axis of the s.c. structure to be normal to these layers.
This direction is called $z$-direction hereafter and the system
occupies the half-space $z>0$. We use the \emph{s-f} model which
is commonly considered as realistic for local-moment structures
\cite{Nagaev,Nolt1,Ovch}. In this model the following Hamiltonian
is used to describe the magnetic semi-infinite crystal:
\begin{equation}
H_t=H_s+H_f+H_{sf}\  ,
\end{equation}
\begin{equation}
H_s=t\sum_{n,n'=1}^{+\infty}\sum_{{\bf r},{\bf r}', \s} |{\bf
r},n;\s\ras\las{\bf r}',n';\s|\  ,
\end{equation}
\begin{equation}\label{Hf}
H_f=-\sum_{n,n'=1}^{+\infty}\sum_{{\bf r},{\bf r}'}J_{{\bf
r}n,{\bf r}'n'} {\bf S}_{{\bf r},n}\cdot{\bf S}_{{\bf r}',n'}\  ,
\end{equation}
\begin{equation}
H_{sf}=-I\sum_{n=1}^{+\infty}\sum_{{\bf r},\s,\s'}|{\bf
r},n;\s\ras ({\bf\s}_{\s\s'}\cdot{\bf S}_{{\bf r},n})\las{\bf
r},n;\s'|\ .
\end{equation}
Here, $H_s$ describes the conduction electrons as $s$-electrons
and gives the electron transfer energy in which $|{\bf
r},n;\s\ras$ is an orbital with spin $\s$ ($=\up,\dd$) at site
${\bf r}$ in the \emph{x-y} plane and layer $n$
$(=1,2,\cdots,+\infty)$. Each lattice point of the semi-infinite
crystal is occupied by a localized magnetic moment, represented
by a spin operator ${\bf S}_{{\bf r},n}$. The direct exchange
coupling between these localized moments is expressed by the
Heisenberg Hamiltonian $H_f$ where $J_{{\bf r}n,{\bf r}'n'}$ is
the exchange coupling constant. The problem with the simple
Heisenberg model in the form (\ref{Hf}) is that the atoms of the
topmost surface layer have fewer neighboring atoms than those of
the bulk; thus, the magnetic order is strongly modified by the
coordination number. This point is considered in calculation of
the layer-dependent magnetization. $H_{sf}$ is the \emph{s-f}
exchange interaction between the $s$-electron and the localized
$f$-spins where ${\bf\s}_{\s\s'}$ is the component of the Pauli
matrix and $I$ is the \emph{s-f} exchange coupling constant. In
this work, we regard a $f$-spin as a classical spin since the
magnitude of the localized spin on a magnetic ion (S=7/2) is
pretty large while the exchange interaction strength $IS$ is kept
finite. The summations in the above equations extend over the
sites of the semi-infinite crystal.

The single-electron Hamiltonian is defined as
\begin{equation}
H=H_{eff}+V\  ,
\end{equation}
where the effective Hamiltonian $H_{eff}$ which describes the
effective medium is expressed as
\begin{equation}\label{H0eff}
H_{eff}=H_s+\sum_{n=1}^{+\infty}\sum_{{\bf r},\s}|{\bf
r},n;\s\ras\ss_{n\s}\las{\bf r},n;\s|\  ,
\end{equation}
and the perturbation term $V$ is written as
\begin{eqnarray}
V&=&H_{sf}-\sum_{n=1}^{+\infty}\sum_{{\bf r},\s}|{\bf
r},n;\s\ras\ss_{n\s}\las{\bf r},n;\s| \\
&=&\sum_{n=1}^{+\infty}\sum_{\bf r}\{\sum_{\s,\s'}|{\bf
r},n;\s\ras[-I({\bf\s}_{\s\s'}\cdot{\bf S}_{{\bf r},n}) \nn \\
&&~~~~~~~~~~~~~~~~~~~~~~~~~~~~~~-\ss_{n\s}\delta_{\s\s'}]\las{\bf
r},n;\s'|\}\\
&=&\sum_{n=1}^{+\infty}\sum_{\bf r}v_{{\bf r},n} \  .
\end{eqnarray}
Here $\ss_{n\s}$ is the layer- and spin-dependent coherent
potential and $v_{{\bf r},n}$ is an isolated potential in site
${\bf r}$ of the $n$-th effective layer.

As in Ref. \cite{Taka96}, we apply the condition that the average
scattering of the $s$-electron by the single $f$-spin embedded in
the effective medium is zero. Thus we define the single-site
$t$-matrix of the \emph{s-f} exchange interaction as
\begin{equation}
t_{{\bf r},n}=v_{{\bf r},n}[1-Pv_{{\bf r},n}]^{-1}
\end{equation}
where $P(\omega)=G^0(\omega-\ss_{n\s})$ is the effective Green's
function and $G^0(\omega)$ is the Green's function of the
nonmagnetic semi-infinite crystal. Here, $t_{{\bf r},n}$ is the
complete scattering associated with the isolated potential
$v_{{\bf r},n}$.

Within the single-site CPA, the condition $\las t_{{\bf
r},n}\ras_{th}=0$ for any ${\bf r}$ in each layer, leads to the
spin-dependent effective medium where an electron is subjected to
a coherent potential, $\ss_{n\up}$ or $\ss_{n\dd}$, according to
the electron spin orientation. The spin-dependent coherent
potentials are energy ($\omega$)-dependent complex potentials and
in this study described by the following equations \cite{Taka96}:
\begin{equation}\label{Sup}
\ss_{n\up}=\frac{F_{n\dd}I^2S(S+1)-IS(1+F_{n\dd}\ss_{n\dd})(A_{n\up}/B_{n\up})}
{1+F_{n\dd}(\ss_{n\dd}-I)-F_{n\dd}IS(A_{n\up}/B_{n\up})}\  ,
\end{equation}
\begin{equation}\label{Sdown}
\ss_{n\dd}=\frac{F_{n\up}I^2S(S+1)+IS(1+F_{n\up}\ss_{n\up})(A_{n\dd}/B_{n\dd})}
{1+F_{n\up}(\ss_{n\up}-I)+F_{n\up}IS(A_{n\dd}/B_{n\dd})}\  ,
\end{equation}
where
\begin{equation}\label{A12}
A_{n\s}=\la\frac{S_n^z/S}{(1-F_{n\s}V_{n\s})
(1-F_{n-\s}U_{n-\s})-F_{n\s}F_{n-\s}W_{n\s}}\ra_{th}\ ,
\end{equation}
\begin{equation}\label{B12}
B_{n\s}=\la\frac{1}{(1-F_{n\s}V_{n\s})
(1-F_{n-\s}U_{n-\s})-F_{n\s}F_{n-\s}W_{n\s}}\ra_{th}\ ,
\end{equation}
\begin{equation}\label{V12}
V_{n\s}=-z_{\s}IS^z_n-\ss_{n\s}\ ,
\end{equation}
\begin{equation}\label{U12}
U_{n\s}=-z_{\s}I(S^z_n-z_{\s})-\ss_{n\s}\ ,
\end{equation}
\begin{equation}\label{W12}
W_{n\s}=I^2[S(S+1)-S^z_n(S^z_n+z_{\s})]\ .
\end{equation}
In above equations $\la\cdots\ra_{th}$ means the thermal average,
$z_{\up}=+1$, $z_{\dd}=-1$, and
\begin{eqnarray}\label{Fn}
F_{n\s}\equiv F_{n\s}(\omega)&=&\frac{1}{N_\parallel}
\sum_{\kp}G^0_{nn}(\kp;\omega-\ss_{n\s})\nn \\
&=&\int_{-\infty}^{\infty} d\epsilon
D^{(2)}(\epsilon)G^0_{nn}(\epsilon;\omega-\ss_{n\s}) \ ,\nn \\
\end{eqnarray}
where $N_\parallel$ is the number of sites per layer, $\kp$ is a
wave vector parallel to the layers, $D^{(2)}(\epsilon)$ is the
density of state for a square lattice with only nearest-neighbor
hopping, and $G^0_{nm}(\kp,\omega)$ is a matrix element of the
Green's function of a semi-infinite crystal which is given in
Appendix A.

It is clear that the Eqs. (\ref{A12})-(\ref{W12}) include only
$S^z_n$ as an $f$-spin operator. Therefore, the thermal average
for fluctuating $f$-spins can be easily calculated using the
molecular-field theory. On the other hand, the existence of
surface modifies the magnetic order in semi-infinite crystals.
Thus, the thermal average of the $f$-spin operator depends on the
layer index $n$. In appendix B we have derived and depicted the
normalized magnetizations $\las S^z_n\ras/S$ which depend on
layer index.

However, with increasing $n$ from the surface layer, the
eigenvalues of semi-infinite Hamiltonian rapidly approach the
bulk Hamiltonian. We find that the localized energies within the
surface layer ($n=1$), or at most, in the second layer ($n=2$),
i.e. the first interior layer, differ from that in the bulk.
Here, we approximate the layer-dependent coherent potential,
$\ss_{n\s}$, by the layer-independent bulk one, $\ss^b_\s$, for
all layers ($n\geq1$). The difference in the coherent potentials
is treated as a local perturbation in the semi-infinite
Hamiltonian and applies only for the layers $n=1$ and $n=2$.
Therefore, we approximate the effective Hamiltonian (\ref{H0eff})
by the following simple form:
\begin{equation}
H_{eff}=H^0_{eff}+U \ ,
\end{equation}
\begin{equation}
H^0_{eff}=H_s+\sum_{n=1}^{+\infty}\sum_{{\bf r},\s}|{\bf
r},n;\s\ras\ss^b_{\s}\las{\bf r},n;\s|\ ,
\end{equation}
\begin{equation}
U=\sum_{n=1}^2\sum_{{\bf r},\s}|{\bf r},n;\s\ras u_{n\s}\las{\bf
r},n;\s| \ ,
\end{equation}
with
\begin{equation}\label{u}
u_{n\s}=\ss_{n\s}-\ss^b_{\s} \ .
\end{equation}
Here, $u_{n\s}$ is a local perturbation for electrons with spin
$\s$ in the $n$-th layer of the semi-infinite crystal.

The layer- and spin-dependent Green's function $\bar{G}_{nm\s}$
which corresponds to $H_{eff}$ is then given by
\begin{eqnarray}\label{gsim}
\bar{G}_{nm\s}(\kp;\omega)&=&\bar{G}^0_{nm\s}(\kp;\omega) \nn \\
&+&\sum_{l=1}^2\bar{G}^0_{nl\s}
(\kp;\omega)u_{l\s}\bar{G}_{lm\s}(\kp;\omega)\ ,\nn \\
\end{eqnarray}
where
\begin{eqnarray}\label{gsim1m}
\bar{G}_{1m\s}&=&\frac{\bar{G}_{1m\s}^0[1-\bar{G}_{22\s}^0u_{2\s}]+\bar{G}_{12\s}^0u_{2\s}
\bar{G}_{2m\s}^0}{[1-\bar{G}_{11\s}^0u_{1\s}][1-\bar{G}_{22\s}^0u_{2\s}]
-\bar{G}_{12\s}^0u_{2\s}\bar{G}^0_{21\s}u_{1\s}}\ ,\nn \\
\end{eqnarray}
\begin{eqnarray}\label{gsim2m}
\bar{G}_{2m\s}&=&\frac{\bar{G}_{2m\s}^0[1-\bar{G}_{11\s}^0u_{1\s}]
+\bar{G}_{21\s}^0u_{1\s}\bar{G}_{1m\s}^0}{[1-\bar{G}_{11\s}^0u_{1\s}][1-\bar{G}_{22\s}^0u_{2\s}]
-\bar{G}_{12\s}^0u_{2\s}\bar{G}^0_{21\s}u_{1\s}}\ ,\nn \\
\end{eqnarray}
and $\bar{G}_{nm\s}^0(\kp;\omega)=G_{nm}^0(\kp;\omega-\ss^b_\s)$
is the corresponding Green's function of $H^0_{eff}$ and can be
obtained from appendix A. Therefore, using Eqs.
(\ref{gsim})-(\ref{gsim2m}) the diagonal elements of the Green's
function for the semi-infinite effective crystal can be resulted
from the relation
\begin{equation}\label{Fbar}
\bar{F}_{n\s}\equiv\bar{F}_{n\s}(\omega)=\int^{\infty}_{-\infty}
d\epsilon D^{(2)}(\epsilon)\bar{G}_{nn\s}(\epsilon;\omega) \ .
\end{equation}

However, before determining $\bar{F}_{n\s}$, we should obtain
$\ss^b_\s$. In Eqs. (\ref{Sup})-(\ref{W12}), by changing
$\ss_{n\s}\rightarrow\ss^b_{\s}$, $B_{n\s}\rightarrow B_{\s}$,
$A_{n\s}\rightarrow A_{\s}$, $V_{n\s}\rightarrow V_{\s}$,
$U_{n\s}\rightarrow U_{\s}$, $W_{n\s}\rightarrow W_{\s}$, and
\begin{eqnarray}
F_{n\s}\rightarrow F^b_{\s}(\omega)&=&\frac{1}{N}\sum_{\bf
k}\frac{1}{\omega-\ss^b_\s(\omega)-\epsilon({\bf k})}\nn\\
&=&\int^{\infty}_{-\infty} d\epsilon D^{(2)}(\epsilon)
\frac{-3i}{W\sqrt{1-9(\frac{\omega-\ss^b_{\s}-\epsilon}{W})^2}}\
,\nn \\
\end{eqnarray}
we can obtain the bulk self-energy \cite{Taka96}. Here $F^b_\s$
is the diagonal element of the bulk effective Green's function,
$N$ is the number of cells in the infinite crystal, and
$\epsilon({\bf k})=2t(\cos k_xa+\cos k_ya+\cos k_za)$.

When $\ss_\s^b$ is determined, we can obtain $\ss_{n\s}$ using
Eqs. (\ref{Sup})-(\ref{W12}). However, in these equations we use
of $\bar{F}_{n\s}$ (Eq. (\ref{Fbar})) instead of $F_{n\s}$ (Eq.
(\ref{Fn})). This procedure is repeated until the calculation
converges. After calculating the layer- and spin-dependent
Green's function $\bar{F}_{n\s}$, the LDOS in layer $n$ can be
determined by the relation
\begin{equation}\label{DOS}
D_{n\s}(\omega)=-\frac{1}{\pi}{\rm Im}
\bar{F}_{n\s}(\omega+i\delta)\ ,
\end{equation}
which should satisfy the following equation in all of the present
numerical calculations
\begin{equation}\label{DOS1}
\int_{-\infty}^{\infty}D_{n\s}(\omega)d\omega=1\ .
\end{equation}
In Eq. (\ref{DOS}), $\delta$ is a positive infinitesimal.

In order to study the electron-spin polarization, we assume that
$N_\up/N_\dd$ is equal to $D_{n\up}(\omega)/D_{n\dd}(\omega)$
where $N_\up$ $(N_\dd)$ is the number of electrons with spin-up
(down). Thus the degree of electron-spin polarization at Fermi
energy $\omega_{_F}$ and in layer $n$ can be given by
\begin{equation}\label{POL}
P_n(\omega_{_F},T)=\frac{D_{n\up}(\omega_{_F})-D_{n\dd}(\omega_{_F})}
{D_{n\up}(\omega_{_F})+D_{n\dd}(\omega_{_F})}\ .
\end{equation}

\section{Results and discussion}
Using the approach described in the previous section, we have
studied the spin-polarized surface states for $IS/W$=0.35 which
corresponds to the intermediate \emph{s-f} exchange interaction.
First we investigate the spin-dependent electronic states in
different positions.

\begin{figure}
\centering \resizebox{0.4\textwidth}{0.35
\textheight}
{\includegraphics{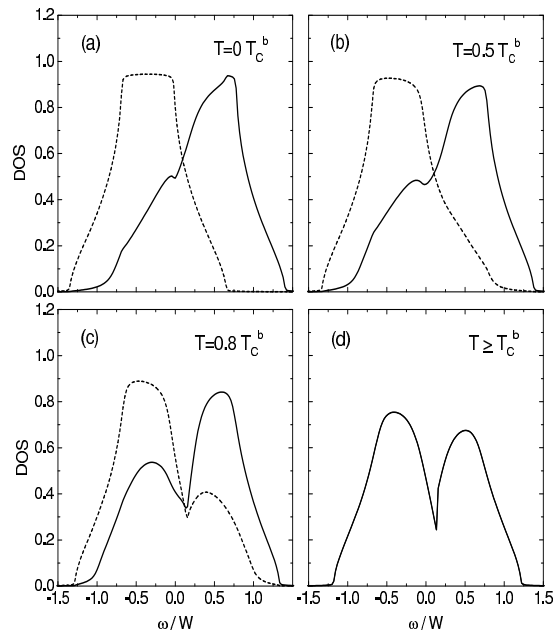}} \caption{The bulk DOS as a function
of energy in various temperatures. The dashed (solid) curve is the
DOS for spin-up (spin-down) electrons.}
\end{figure}

In Fig. 1, we have shown the DOS as a function of energy for a
s.c. crystal at $T$=0, 0.5, 0.8 $T_c^b$, and $T\geq T_c^b$. It is
clear that the largest spin-splitting between both spin-up and
spin-down bands occurs at zero temperature. At this temperature,
the spin-up DOS is similar to the DOS of a nonmagnetic crystal.
Since at zero temperature (completely ferromagnetic case) all the
local moments are aligned at $z$-direction, the magnetic system is
periodic. As a result, the spin-up states are only shifted to the
low energy side with no damping, and diminish at high energies
where the coupling of the states are mainly antiparallel. On the
other hand, the spin-down electrons can be flipped; thus, the
spin-down band is shifted to the high energy region, while the
bottom of the spin-down band extends down. With increasing the
temperature, the local moments fluctuate and the spin-splitting
between both spin bands decreases. At $T\geq T_c^b$, the
spontaneous magnetization is zero and hence the DOS are
spin-independent.

\begin{figure}
\centering \resizebox{0.4\textwidth}{0.45\textheight}
{\includegraphics{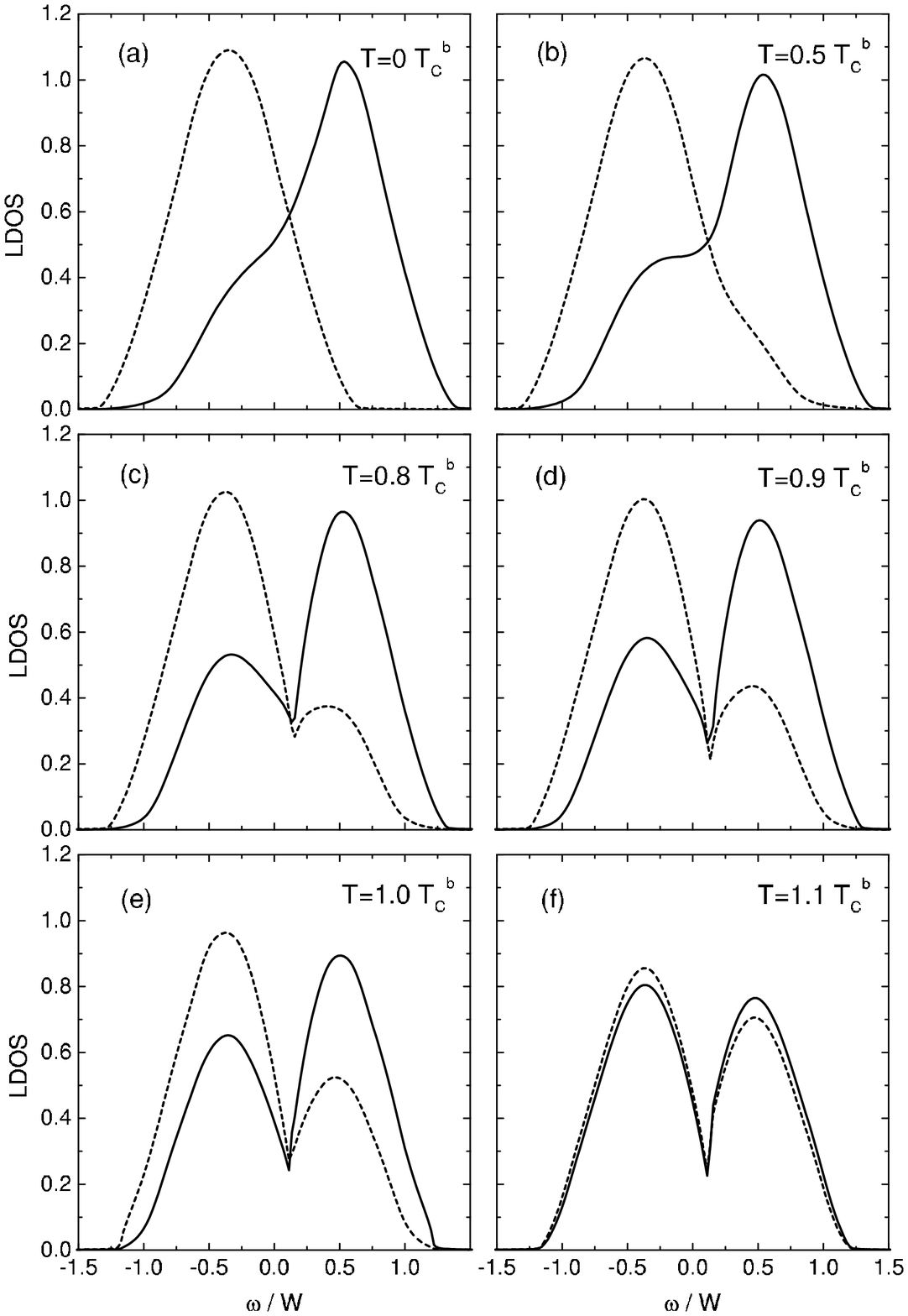}} \caption{The LDOS for the surface
layer ($n$=1) as a function of energy in various temperatures. The
dashed (solid) curve is the LDOS for spin-up (spin-down)
electrons. Here, $J_{\parallel}$=1.50 $J$ and $J_{\perp}$=1.15
$J$.\\}

\centering \resizebox{0.4\textwidth}{0.45\textheight}
{\includegraphics{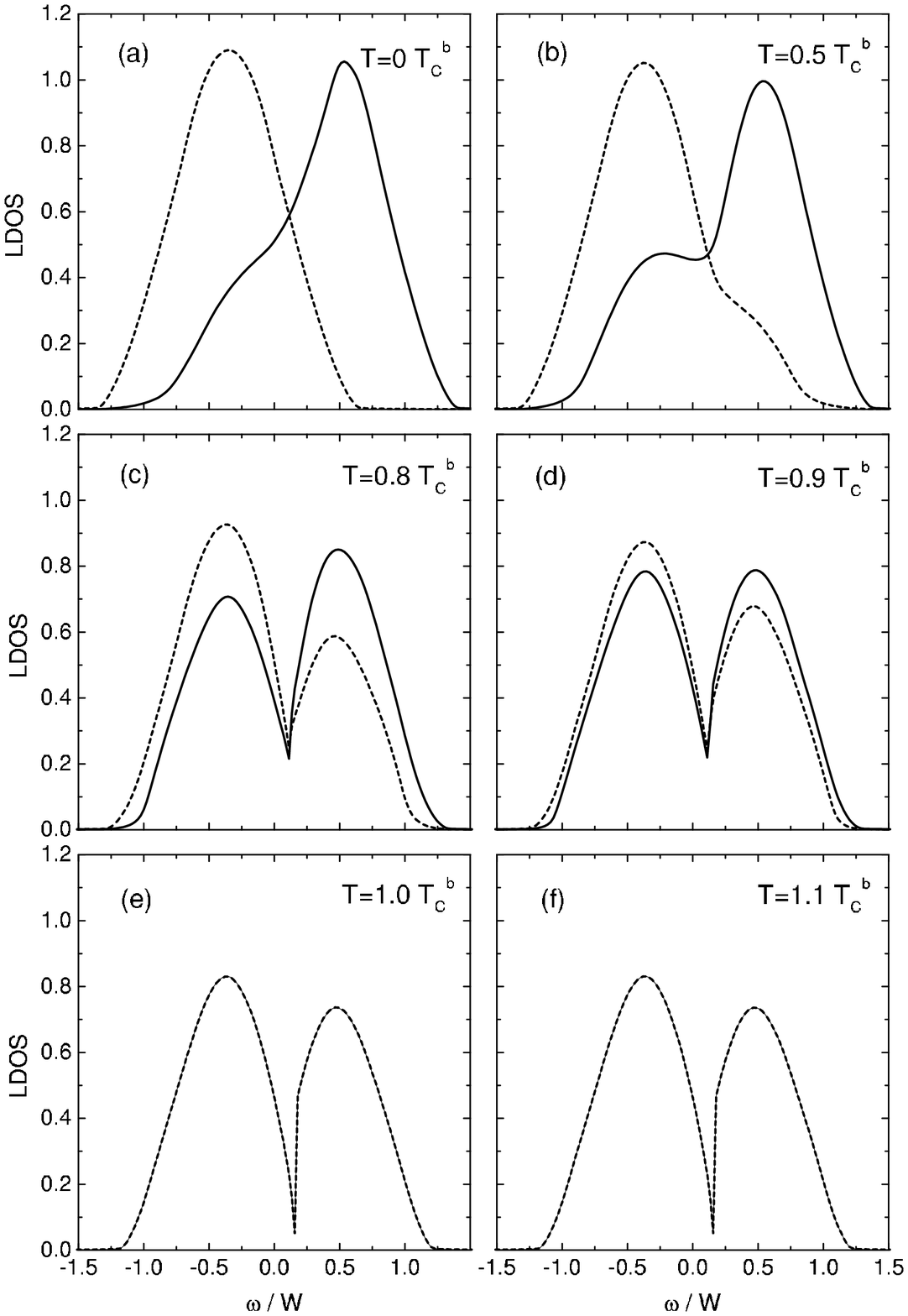}} \caption{The LDOS for the surface
layer ($n$=1) as a function of energy in various temperatures. The
dashed (solid) curve is the LDOS for spin-up (spin-down)
electrons. Here, $J_{\parallel}$=0.80 $J$ and $J_{\perp}$=0.95
$J$.}
\end{figure}

In Figs. 2 and 3, we have depicted the surface LDOS for spin-up
and spin-down electrons at different temperatures. The curves in
Fig. 2 correspond to the case in which the surface orders at a
temperature $T_C^s$ (surface transition) larger than the bulk
transition temperature $T_C^b$. In this regard, we have chosen
$J_{\parallel}$=1.5 $J$ and $J_{\perp}$=1.15 $J$. The curves in
Fig. 3, however correspond to the case in which both the bulk and
the surface order at identical temperature (ordinary transition).
In such case we have assumed that $J_{\parallel}$=0.8 $J$ and
$J_{\perp}$=0.95 $J$. If we consider the layer-dependent
magnetization in Fig. 8, we will see that, at low temperatures,
the LDOS in the case of surface transition is similar to those
obtained in the case of ordinary transition. At zero temperature,
the spin-up LDOS is similar to the surface LDOS in a nonmagnetic
semi-infinite crystal \cite{Kalkstein}. With increasing
temperature from zero to $T\simeq$0.5 $T_C^b$, the LDOS does not
change considerably in both cases.  The spin-splitting in the LDOS
is a measure of the electron-spin polarization which depends on
the layer index. Further increase of temperature, decreases the
spin-splitting of the LDOS. Fig. 2 shows that, for the surface
transition the spin-splitting is not zero at $T=T_C^b$ and even at
$T=$1.1 $T_C^b$, however for the ordinary transition the
spin-splitting will be zero at $T=T_C^b$ as shown in Fig. 3.

\begin{figure}
\centering \resizebox{0.4\textwidth}{0.45\textheight}
{\includegraphics{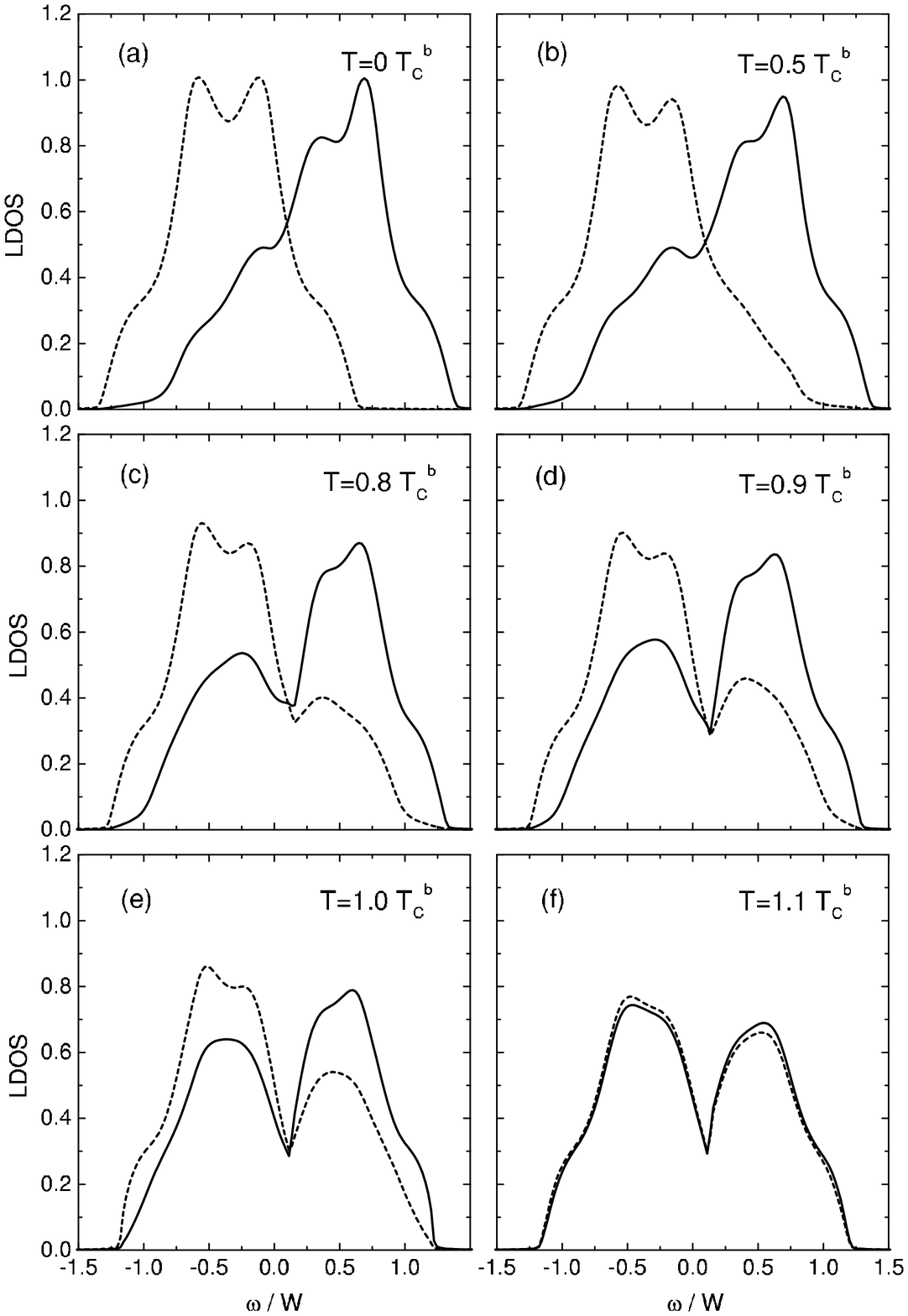}} \caption{The LDOS for the first
interior layer ($n$=2) as a function of energy in various
temperatures. The dashed (solid) curve is the LDOS for spin-up
(spin-down) electrons. Here, $J_{\parallel}$=1.50 $J$ and
$J_{\perp}$=1.15 $J$.\\}

\centering \resizebox{0.4\textwidth}{0.45\textheight}
{\includegraphics{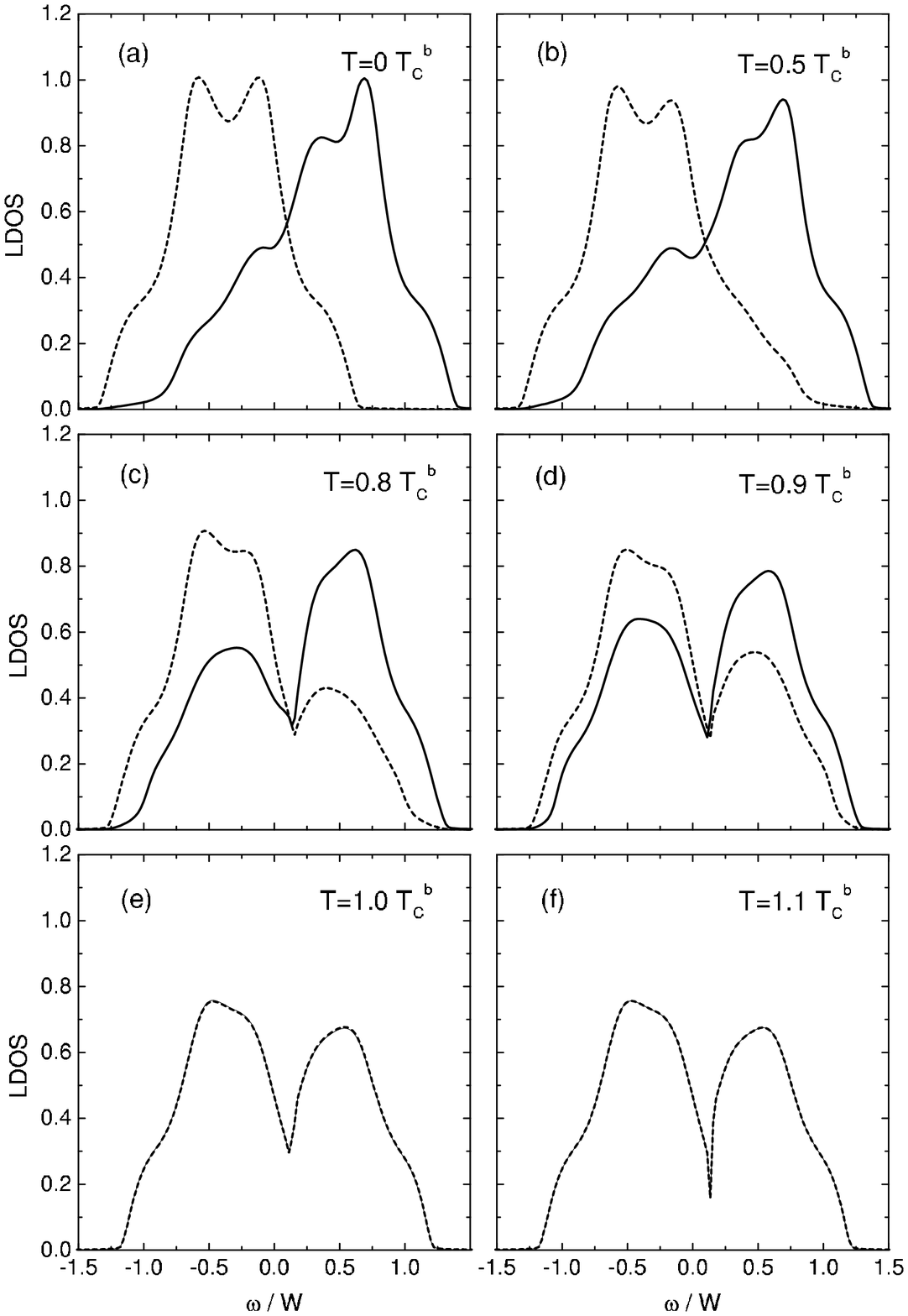}} \caption{The LDOS for the first
interior layer ($n$=2) as a function of energy in various
temperatures. The dashed (solid) curve is the LDOS for spin-up
(spin-down) electrons. Here, $J_{\parallel}$=0.80 $J$ and
$J_{\perp}$=0.95 $J$.}
\end{figure}

In Figs. 4 and 5 we have depicted the temperature and spin
dependence of LDOS in the second layer at the surface and
ordinary transition cases, respectively. The electronic states
slightly resemble the DOS function for the infinite crystal. In
this layer the spin-down band, as the previous electronic states,
has a tail which reaches the edge of the spin-up band even at low
temperatures. As the temperature increases, the spin-up (-down)
LDOS shifts to high (low) energy side; so that, at paramagnetic
region the LDOS for both spin-up and spin-down bands completely
coincide to each other. It is clear that in all cases, there is a
cusp in the LDOS which is due to the $s$-$f$ exchange coupling
strength.

\begin{figure}
\resizebox{0.4\textwidth}{0.6\textheight}
{\includegraphics{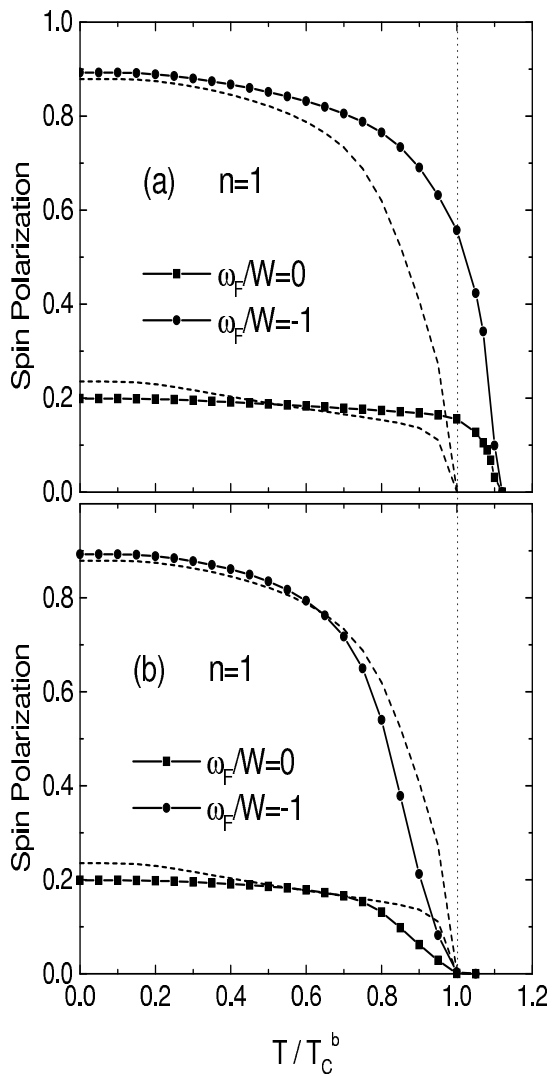}} \caption{The degree of spin
polarization for the surface layer ($n$=1) as a function of
normalized temperature $T/T^b_C$: (a) $J_{\parallel}$=1.50 $J$ and
$J_{\perp}$=1.15 $J$, (b) $J_{\parallel}$=0.80 $J$ and
$J_{\perp}$=0.95 $J$. The dashed curve is the bulk electron-spin
polarization.}
\end{figure}

If we compare the Figs. 2(f) and 4(f) (at $T$=1.1 $T_C^b$), we
will find that the value of spin-splitting at the surface layer
is slightly more than the first interior layer (second layer).
Therefore, as one proceeds into the crystal, the value of spin
polarization begins to decrease at temperatures higher than
$T_C^b$. When we are far enough from the surface, the surface
effect on the spin polarization completely disappears and at
$T\geq T_C^b$ the electron-spin polarization becomes zero.

\begin{figure}
\centering \resizebox{0.4\textwidth}{0.6\textheight}
{\includegraphics{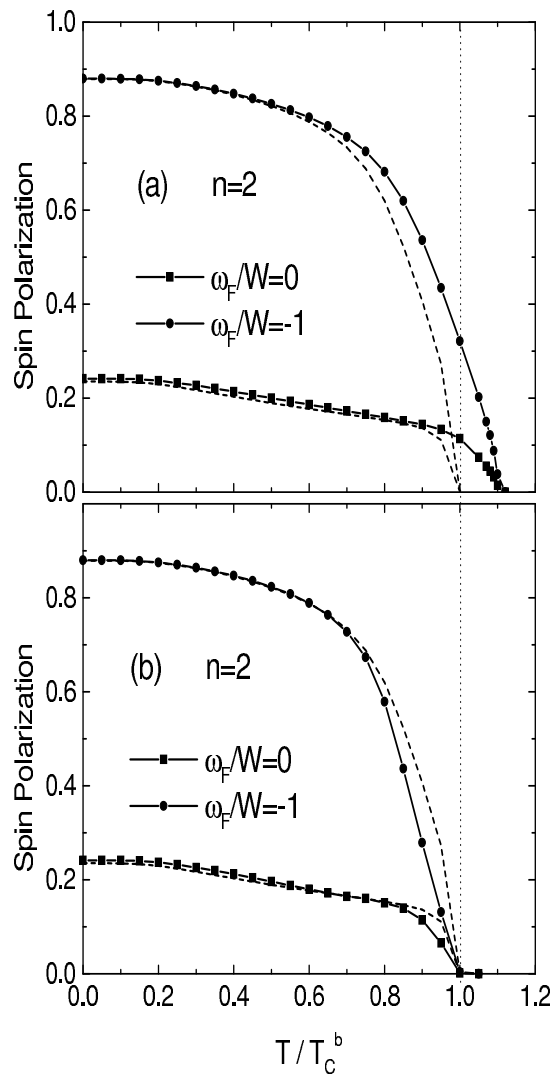}} \caption{The degree of spin
polarization for the first interior layer ($n$=2) as a function of
normalized temperature $T/T^b_C$: (a) $J_{\parallel}$=1.50 $J$ and
$J_{\perp}$=1.15 $J$, (b) $J_{\parallel}$=0.80 $J$ and
$J_{\perp}$=0.95 $J$. The dashed curve is the bulk electron-spin
polarization.}
\end{figure}

In order to gain further insight into the surface effects on the
LDOS, we have shown in Figs. 6 and 7, the temperature dependence
of the electron-spin polarization in both surface and ordinary
transition cases for the first and second layers, respectively. In
most of the magnetic phenomena at surfaces, interfaces and in thin
films, such as spin-polarized electron emission and spin-polarized
transport through a magnetic film, the electrons at the Fermi
surface play an essential role. Therefore, the electron-spin
polarization depends on the position of Fermi energy. Here, in
both Fig. 6 and 7, we have calculated the spin polarization at
$\omega_{_F}/W=-1$ (near the band edge) and at $\omega_{_F}/W=0$
(near the band center). For comparison, we have shown the
electron-spin polarization for the infinite crystal. It is clear
that, in surface transition case when $\omega_{_F}/W=-1$, the spin
polarization on the surface ($n$=1) and in the first interior
layer ($n$=2) is higher than the bulk one. The spin polarization
on the surface layer differs with the bulk one at all
temperatures. This difference near $T_C^b$ reaches its maximum
value. However, due to the reduction of the spin polarization in
the first interior layer in comparison with the surface layer, the
difference in the spin polarization of this layer and the bulk
appears above 0.6 $T_C^b$.

The difference of the spin polarizations between both layers and
the bulk is originated from the strong exchange coupling with
magnetic ions. At $\omega_{_F}/W=0$, the spin polarizations have
low values at all temperatures, because the difference between two
spin bands is small. If we consider the ordinary transition case,
we will see that, near $T_C^b$ that the surface magnetization has
the most difference with the bulk one, the bulk spin polarization
is more than the value of spin polarization at the layers $n=1$
and $n=2$.

We applied the present model for s.c. local moment systems such as
Eu chalcogenides. This model is also applicable to the case of hcp
Gd. However, for an atom at the Gd(0001) surface, we have six
nearest-neighbor atoms at the surface, and three nearest-neighbor
atoms in the second layer. Applying the molecular-field theory to
the case of a hcp(0001) surface, we find that, for an exchange
coupling ratio that satisfies $J^s\geq\frac{4}{3}J$, there is a
surface transition \cite{Shick}. Here, $J^s$ is the exchange
coupling constant at the surface.

It should be noted that, real surfaces are usually not perfectly
smooth and display some degree of roughness due to the presence of
surface defects, such as for example steps, islands, or vacancies.
Impurities are also often encountered at crystalline surfaces and
may be viewed as the source of some disorder at the surface. All
these defects have some impact on magnetic surface quantities.
Based on the single-site CPA, Hong \cite{Hong} studied the LDOS
near a rough surface. The numerical results show that, the LDOS
depend sensitively on surface roughness. On the other hand, Zhao
et al. \cite{Zhao} using Ising-model Monte Carlo simulations have
shown that surface roughness strongly affects surface
magnetization, not only changing the surface magnetic ordering
temperature but also modifying the magnitude of magnetization in a
complex fashion. Therefore, both the LDOS and the surface
magnetization are sensitive to the surface defects and
consequently the resulting spin polarization is strongly
influenced by the vacancies, islands, steps and the other surface
defects.

The above results show that the exchange coupling between the
surface magnetic ions plays an important role in the degree of
spin polarization for the emitted electrons from the free surface
or the transmitted electrons through interfaces. Hence, in
designing the spin electronic or ``spintronic" devices the
information about the surface effects on the degree of
spin-polarization plays an essential role in generation of highly
polarized electrons.

\section{Conclusions}
In summary, we have presented a new formalism to study the
spin-polarized surface states in the local moment systems. Based
on the single-site CPA for the \emph{s-f} model and the mean-field
theory for the layer-dependent magnetization we have investigated
how the surface influences the spin-dependent electronic states.
The effect of the semi-infinity of the crystal was taken into
account by a localized perturbation method. Assuming a s.c. band
structure for the $s$ electrons we obtained the numerical results
for the LDOS and the electron-spin polarization for various
temperatures and in two different behaviors at the surface;
ordinary and surface transitions. These transitions depend on the
values of the exchange coupling constants between atoms on the
surface and between atoms in the layers $n=1$ and $n=2$. The
results showed that in the surface transition case, in which the
coupling constants are higher than the corresponding in the bulk,
the surface Curie temperature $T_C^s$ lies well above the bulk
Curie temperature. Therefore, at $T_C^b<T<T_C^s$ the electron-spin
polarization on the surface has a finite value, while the bulk
spin polarization is zero. The existence of the surface
magnetization and hence, the spin-polarized surface states above
the bulk Curie temperature can be useful for development of new
electron-polarized devices.

\section*{Acknowledgments}
This work was supported by the Payame Noor University.

\appendix
\section{Green's function of a semi-infinite crystal}
The existence of surface affects the electronic eigenfunction;
therefore, due to the surface effect, the Green's function of a
semi-infinite crystal is essentially layer-dependent. In this
regard, for the eigenfunction of a semi-infinite crystal we use
of a simple form as:
\begin{equation}\label{A1}
|\kp,k_\perp\ras=\frac{1}{\sqrt{N_\perp}}\sum_n\varphi_n(k_\perp)|\kp,n\ras\
,
\end{equation}
where the layer-dependent coefficient $\varphi_n(k_\perp)$ is
given by the standing wave
\begin{equation}\label{A2}
\varphi_n(k_\perp)=\sqrt{2} \sin(k_\perp na)\  ,
\end{equation}
and satisfies the following orthogonality conditions
\begin{equation}\label{A3}
\sum_n\varphi_n^*(k_\perp)\varphi_n(k_\perp')=\delta_{k_\perp,k_\perp'}\
,
\end{equation}
\begin{equation}\label{A4}
\sum_{k_{\perp}}\varphi_n^*(k_\perp)\varphi_{n'}(k_\perp)=\delta_{n,n'}\
.
\end{equation}

Using Eqs. (\ref{A1}) and (\ref{A2}), we obtain the
layer-dependent Green's function for a semi-infinite crystal with
a (001) surface as follows;
\begin{equation}\label{A5}
G^0_{nm}(\kp;\omega)=\frac{1}{N_\perp}\sum_{k_{\perp}}\frac{\varphi_n^*(k_\perp)
\varphi_{m}(k_\perp)}{\omega-\epsilon_\kp-2t\cos(k_\perp a)}\ .
\end{equation}
Here $N_\perp$ is the number of layers parallel to the surface and
$\epsilon_\kp=2t(\cos k_x a+\cos k_y a)$.

By transforming the sum over $k_\perp$ in Eq. (\ref{A5}) to an
integral over $k_\perp a=\theta$, the Green's function in a mixed
Bloch-Wannier representation for a semi-infinite crystal is then
given as
\begin{equation}
G_{nm}^0(\kp;\omega)=\frac{-3i}{W\sqrt{1-\eta^2}}[f(\eta)^{|n-m|}-f(\eta)^{|n+m|}]\
,
\end{equation}
where $W=6t$ is the half-bandwidth and
\begin{equation}
\eta=3(\omega-\epsilon_{\kp})/W\  ,
\end{equation}
\begin{equation}
f(\eta)=\eta-i\sqrt{1-\eta^2}\ .
\end{equation}
\section{Layer-dependent magnetization}
Here, we like to present a generalized molecular-field theory to
calculate the layer-dependent magnetization for a magnetic
semi-infinite crystal. Due to a wide range of physical effects
near surfaces, surface properties may be significantly different
from bulk properties \cite{Rau88,Rau86,Tang,Cottam}. These are
partly taken into account by a deviation of exchange coupling
constant $J_{{\bf r}n,{\bf r'}n'}$ in nearest neighbor
approximation and in the neighborhood of surface. For simplicity,
we assume in all calculations that only the coupling constants of
the local magnetic moments at the surface, $J_{{\bf r}1,{\bf
r'}1}=J_\|$, and also between atoms in the surface and second
layer, $J_{{\bf r}1,{\bf r}2}=J_\bot$, differ from the bulk
exchange coupling constant $J_{{\bf r}n,{\bf r'}n'}=J$.

\begin{figure}
\centering \resizebox{0.4\textwidth}{0.6\textheight}
{\includegraphics{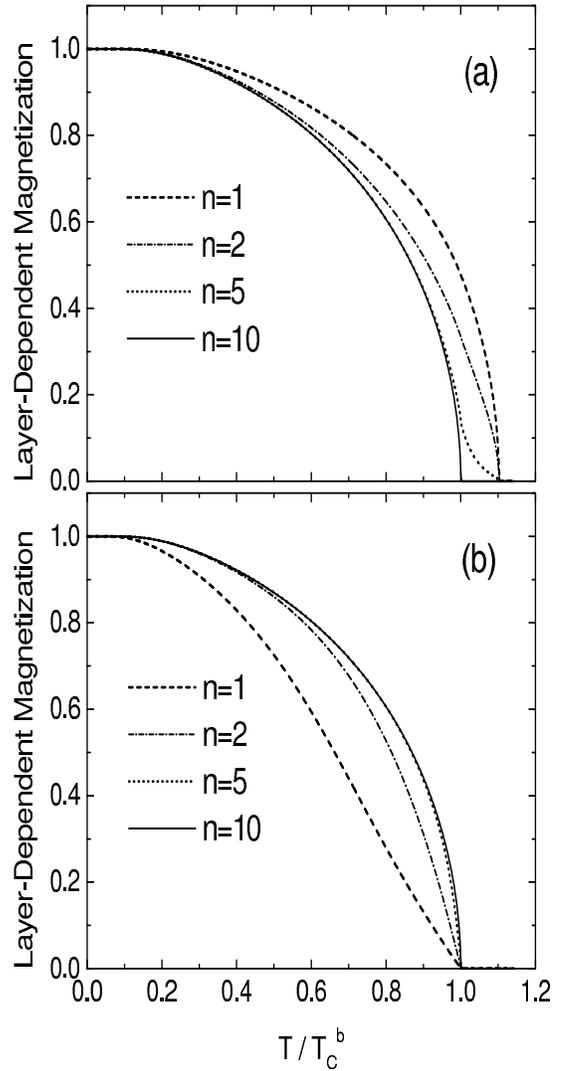}} \caption{Layer-dependent normalized
magnetization, $\las S_n^z\ras/S$, as a function of normalized
temperature, $T/T^b_C$. (a) $J_{\parallel}$=1.50 $J$ and
$J_{\perp}$=1.15 $J$, (b) $J_{\parallel}$=0.80 $J$ and
$J_{\perp}$=0.95 $J$.}
\end{figure}

The molecular-field equations given by Cottam \cite{Cottam} for
spontaneous magnetizations of the semi-infinite crystals with s.c.
structure and (001) orientation of the surface, are of the form
\begin{eqnarray}
\frac{\las{S_1^z}\ras}{S}&=&B_s\left[\frac{S}{2\tau(S+1)}
\left(4\frac{J_{\parallel}}{J}\frac{\las{S_1^z}\ras}{S}+
\frac{J_{\perp}}{J}\frac{\las{S_2^z}\ras}{S}\right)\right]\label{e5}\
,\nn \\
&&\\
\frac{\las{S_2^z}\ras}{S}&=&B_s\left[\frac{S}{2\tau(S+1)}
\left(\frac{J_{\perp}}{J}\frac{\las{S_1^z}\ras}{S}+
4\frac{\las{S_2^z}\ras}{S}+\frac{\las{S_3^z}\ras}{S}\right)\right]
\label{e6}\ ,\nn \\
&&\\
\frac{\las{S_n^z}\ras}{S}&=&B_s\left[\frac{S}{2\tau(S+1)}
\left(\frac{\las{S_{n-1}^z}\ras}{S}+4\frac{\las{S_n^z}\ras}{S}+
\frac{\las{S_{n+1}^z}\ras}{S}\right)\right] \nn \\
&& ~~~~~~~~~~~~~~~~~~~~~~~~~~~~~~~~~~~~~~~~~~~~~ n\geq3\ ,
\label{e7}
\end{eqnarray}
where $B_s(x)$ is the Brillouin function and $\tau=T/T^b_C$ in
which $T^b_C$ is the bulk Curie temperature of the sample.
Depending on the values of $J_{\parallel}$ and $J_{\perp}$, two
different behaviors at the surface can be obtained: (i) If the
coupling constants $J_{\parallel}$ and $J_{\perp}$ are larger than
$J$, the magnetization at the surface and even in the layers near
to the surface may have nonzero values at $T>T^b_C$. Therefore,
these layers are in the ferromagnetic phase for $T^b_C<T<T_C^s$
($T_C^s$ is the surface Curie temperature), while the bulk is in
paramagnetic state. (ii) If $J_{\parallel}$ and $J_{\perp}$ are
smaller than $J$, the surface and the layers near the surface may
have weaker magnetization than the bulk at temperatures near
$T^b_C$. The normalized magnetization in each case can be
determined using the self-consistent solution of Eqs.
(\ref{e5})-(\ref{e7}). The layer-dependent magnetization for both
cases has been shown in Fig. 8. It is important to note that, the
temperature dependence of the magnetizations depends strongly on
the boundary conditions imposed in the calculation. Here, the bulk
boundary conditions were imposed in 10th layer \cite{Moran}.


\begin{thebibliography}{plane}

\bibitem{Labella}
{V.P. LaBella, D.W. Bullock, Z. Ding, C. Emery, A. Venkatesan,
W.F. Oliver, G.J. Salamo, P.M. Thibado, and M. Mortazavi, Science
{\bf 292}, 1518 (2001).}

\bibitem{Koon}
{N.C. Koon, B.T. Jonker, F.A. Volkening, J.J. Krebs, and G.A.
Prinz, Phys. Rev. Lett. {\bf 59}, 2463 (1987).}

\bibitem{Stam}
{M. Stamspanoni, A. Vaterlaus, M. Aeschlimann, and F. Meier,
Phys. Rev. Lett. {\bf 59}, 2483 (1987).}

\bibitem{Liu1}
{C. Liu, E.R. Mong, and S.D. Bader, Phys. Rev. Lett. {\bf 60},
2422 (1988).}

\bibitem{Durr}
{W. Durr, M. Taborelli, O. Paul, R. Germar, W. Gudot, D. Pescia,
and M. Landolt, Phys. Rev. Lett. {\bf 62}, 206 (1989).}

\bibitem{Liu2}
{C. Liu and S.D. Bader, J. Appl. Phys. {\bf 67}, 5758 (1990).}

\bibitem{Celotta}
{R.J. Celotta, D.T. Pierce, G.C. Wang, S.D. Bader, and G.P.
Felcher, Phys. Rev. Lett. {\bf 43}, 728 (1979).}

\bibitem{Alvarado}
{S.F. Alvarado, M. Campagna, and H. Hopster, Phys. Rev. Lett.
{\bf 48}, 51 (1982).}

\bibitem{Weller85}
{D. Weller, S.F. Alvarado, W. Gudat, K. Schroder and M. Capagna,
Phys. Rev. Lett. {\bf 54}, 1555 (1985).}

\bibitem{Rau87}
{C. Rau, M. Robert, Phys. Rev. Lett. {\bf 58}, 2714 (1987).}

\bibitem{Rau88}
{C. Rau, C. Jin, and M. Robert, J. Appl. Phys. {\bf 63}, 3667
(1988).}

\bibitem{Mary}
{M. Marynowski, W. Franzen, M. El-Batanouny, V. Staemmler, Phys.
Rev. B {\bf 60}, 6053 (1999).}

\bibitem{Bata}
{M. El-Batanouny, J. Phys.: Condens. Matter {\bf 14}, 6281
(2002).}

\bibitem{Donath}
{M. Donath, B. Gubanka, F. Passek, Phys. Rev. Lett. {\bf 77}, 5138
(1996).}

\bibitem{Arnold}
{C.S. Arnold, D.P. Pappas, Phys. Rev. Lett. {\bf 85}, 5202
(2000).}

\bibitem{Kurz}
{Ph. Kurz, G. Bihlmayer, and S. Bl\"ugel, J. Phys.: Condens.
Matter {\bf 14}, 6353 (2002).}

\bibitem{Schiller1}
{R. Schiller, W. M\"uller, and W. Nolting, J. Phys.: Condens.
Matter {\bf 11}, 9589 (1999).}

\bibitem{Schiller2}
{R. Schiller and W. Nolting, Phys. Rev. B {\bf 60}, 462 (1999).}

\bibitem{Nagaev}
{E.L. Nagaev, Phys. Status Solidi B {\bf 65}, 11 (1974).}

\bibitem{Taka96}
{M. Takahashi and K. Mitsui, Phys. Rev. B {\bf 54}, 11298 (1996).}

\bibitem{Nolt1}
{W. Nolting, Phys. Status Solidi B {\bf 96}, 11 (1979).}

\bibitem{Ovch}
{S.G. Ovchinikov, Phase Transitions {\bf 36}, 15 (1991).}

\bibitem{Kalkstein}
{D. Kalkstein and P. Soven, Surf. Sci. {\bf 26}, 85 (1971).}

\bibitem{Shick}
{A.B. Shick, W.E. Pickett, C.S. Fadley, Phys. Rev. B {\bf 61}
R9213 (2000).}

\bibitem{Hong}
{K.M. Hong, Solid State Commun. {\bf 66} 241 (1988).}

\bibitem{Zhao}
{D. Zhao, F. Liu, D.L. Huber, M.G. Lagally, Phys. Rev. B {\bf 62}
11316 (2000).}

\bibitem{Rau86}
{C. Rau and S. Eichner, Phys. Rev. B {\bf 34}, 6347 (1986).}

\bibitem{Tang}
{H. Tang, D. Weller, T.G. Walker, J.C. Scott, C. Chappert, H.
Hopster, A.W. Pang, D.S. Dessau, and D.P. Pappas, Phys. Rev.
Lett. {\bf 71}, 444 (1993).}

\bibitem{Cottam}
{M.G. Cottam and D.R. Tilly, {\it Introduction to surface and
superlattice excitations}, Cambrdige, New Rochelle Melburne,
Sydney, Cambridge University Press 1989.}

\bibitem{Moran}
{J.L. Mor\'an-L\'opez and J.M. Sanchez, Phys. Rev. B {\bf 39},
9746 (1989).}
\end{thebibliography}
\end{document}